\documentclass[twocolumn,prl,showpacs]{revtex4}
\usepackage{graphicx}

\def\<{\langle}
\def\>{\rangle}

\def\half{{\frac{1}{2}}}
\def\be{\begin{equation}}
\def\ee{\end{equation}}
\def\Sent{S_{\rm ent}}
\def\delS{\delta S_{\rm imp}}

\begin{document}
\preprint{cond-mat} \title{Entanglement entropy in a boundary impurity model}

\author{Gregory Levine}

\address{Department of Physics and Astronomy, Hofstra University,
Hempstead, NY 11549}

\date{\today}

\begin{abstract}
Boundary impurities are known to dramatically alter certain bulk properties of $1+1$ dimensional strongly correlated systems. The entanglement entropy of a zero temperature Luttinger liquid bisected by a single impurity is computed using a novel finite size scaling/bosonization scheme.    For a Luttinger liquid of length $2L$ and UV cut off $\epsilon$,  the boundary impurity correction ($\delS$) to the bulk logarithmic entanglement entropy ($S_{\rm ent} \propto \ln{L/\epsilon}$) scales as $\delS \sim y_r \ln{L/\epsilon}$, where $y_r$ is the renormalized backscattering coupling constant.  In this way, bulk entanglement entropy within a region is related to scattering through the region's boundary.  In the repulsive case ($g<1$),  $\delS$ diverges (negatively) suggesting that the bulk entropy vanishes.   Our results are consistent with the recent conjecture that entanglement entropy decreases irreversibly along renormalization group flow.   

\end{abstract}

\maketitle

Quantum field theories describe coupled quantum oscillators and, therefore, even the ground state of noninteracting fields may exhibit quantum correlations over a long range.  Consider the ground state of the free electromagnetic field.  An electric field amplitude $E$ at a point $x$ is quantum correlated with an amplitude $E^\prime$ at a point $x^\prime$ in that there are contributions to the ground state wavefunction containing such products as $\ldots|E\>_x|E^\prime\>_{x^\prime}\ldots$.  Since the ground state contains superpositions of such states, it is said to be {\sl entangled}---even though it is a free field. If, in a $d$-dimensional spacetime, a spatial box of volume $L^{d-1}$ is formed,  it follows that the degrees of freedom which reside exclusively in the box will appear to be in a mixed state.  The degree of mixing may be characterized by the von Neumann entropy, $S = -{\rm tr}\rho \ln{\rho}$, where  the reduced density matrix $\rho$ has been formed by tracing over the degrees of freedom exterior to the volume $L^{d-1}$.

Geometric or {\sl entanglement} entropy formed in this fashion was introduced in the context of black hole quantum mechanics and Hawking-Bekenstein entropy \cite{BirrellDavies}, where it was found that entanglement entropy is not an extensive quantity but, rather,  scales as the area of the bounding surface, $S \propto L^{d-2}$ \cite{Srednicki}.  This highly suggestive result is believed to bear some relation to holographic principle proposals \cite{Bousso}. In condensed matter physics, the role of entanglement entropy in understanding quantum critical phenomena \cite{Vidal} and quantum phase transitions \cite{QPT} has recently been emphasized.  $1+1$ dimensional conformal field theories---which describe critical spin chains, Luttinger liquids and other massless theories---have pointlike  bounding surfaces;  remarkably, the entanglement entropy was shown to depend universally upon the central charge of the theory and to diverge logarithmically with the length of the subsystem \cite{CallanWilczek,Holzhey,PreskillStrominger,Cardy_rev}.  Specifically, the entropy is given by  $S =  \frac{c}{3} \ln{L/\epsilon}$ where $c$ is the central charge. 

Studies of entanglement entropy have, so far, been restricted to homogeneous models. [For a  recent exception, see \cite{Refael}.] However, consider the artificial introduction of an inhomogeneity at one boundary point.  Quantum entanglements across the boundary must be associated with {\sl scattering} through the boundary and, therefore,  a boundary impurity that alters the bulk  conductance is expected to alter the bulk entanglement entropy.  Specifically, backscattering from an impurity should be associated with negative corrections to the bulk entanglement entropy---if a sensible perturbative approach to this highly non-local quantity can be found.   In this manuscript, we develop a scheme by which these corrections to entropy may be computed for a general boundary impurity model and demonstrate a connection between the scaling of backscattering and that of entanglement entropy for the specific model of an impurity in a Luttinger liquid (LL). Significantly, we find a logarithmic divergence in the entropy computed in this fashion which has its origin in an infrared divergence in the effective boundary theory. Finally, our  perturbation theory directly addresses and supports the conjecture \cite{Vidal} that entanglement entropy decreases irreversibly along renormalization group flows.

We adopt the model of Kane and Fisher (KF) \cite{KaneFisher}, in which a single impurity in a Luttinger liquid  was shown to have a dramatic effect---either effectively decoupling the two sides, or effectively vanishing, depending on whether the LL is repulsive ($g<1$) or attractive ($g>1$).  The behavior of $\Sent$ suggests the following puzzle: If one considers (following \cite{PreskillStrominger}) a "mixed" real space/momentum space basis where at wavenumber $k$  there are a series of ($kL$) spatial modes of width $k^{-1}$, entanglement entropy arises from those spatial modes which "straddle" the boundary; counting them naively gives $\ln{(k_{\rm max}/k_{\rm min})} \sim \ln{L/\epsilon}$ \cite{basis_explanation}.  The UV divergence of the entropy is a consequence of the unlimited number modes close the boundary.  If an impurity is introduced into a repulsive LL, the effective strength grows with inverse wavenumber and therefore the contributions to entanglement become weak at the longest wavelengths.  Using the KF scaling of impurity strength, one might estimate the correction to entanglement entropy, $\delS$, by excluding contributions to the sum above from wavenumbers less than $V_0 (L/\epsilon)^{1-g}$, the approximate energy scale of the impurity in a finite lattice ($1/T \gg L$). This estimate gives $\delS \sim (g-1) \ln{(L/\epsilon)} - {\rm const}$ which demonstrates the expected reduction.  However, this estimate suggests that $S_{\rm ent}$ may still diverge as $L/\epsilon \rightarrow \infty$, contrary to the expectation that the impurity effectively "disconnects" the two sides for $g<1$. 

To proceed, $S_{\rm ent}$ must be directly calculated or expressed as operators with known LL scaling dimensions. However, $S_{\rm ent}$ is an unusual, non-local quantity and despite its superficial resemblance to a log-scaling marginal operator, it has no simple operator representation within the LL.  An elegant geometric approach to this problem was pioneered by  Callan and Wilczek \cite{CallanWilczek} and Cardy \cite{CardyPeschel}, who introduced a euclidean path integral representation for the geometric entropy of a $1+1$ dimensional field theory computed on the half-infinite spatial domain.  First expressing the ground state wavefunctional as a euclidean path integral
\begin{equation}
\Psi[\phi(x)] =\<\phi|0\> \propto \int{D\phi e^{-S[\phi]}},
\end{equation}
the ground state density matrix functional may be written as a double path integral where one identifies the imaginary time domain to be $\tau \in (-\infty,0)$ for one path and $\tau \in (\infty,0)$ for the other.  Now the reduced density matrix functional for the fields $\phi_\pm$ on halfline ($x>0$), $\langle \phi_+ | \rho | \phi_- \rangle$, may be expressed as a path integral over the {\sl entire} $1+1$ dimensional euclidean space ($x,\tau \in (-\infty,\infty)$) with $\phi_\pm $ as boundary conditions on either side of a cut along the polar angle $\theta=0$ (figure 1).
If this cut is enlarged to a finite "deficit angle," $\alpha$, the path integral is now evaluated on a cone.
\begin{equation}
\label{partition}
Z_\alpha \equiv \langle \phi_+ | \rho | \phi_- \rangle =  \int_{R^2/\{\theta=0\}}{D\phi e^{-S[\phi]}}
\end{equation}
Using the replica trick, it was shown that the entropy $\Sent = -{\rm tr}\rho \ln{\rho}$ may be written as a variation of the partition function, $Z_\alpha$, with respect to $\alpha$ \cite{CallanWilczek}.  Specifically, $\Sent = (-2\pi {d \over {d\alpha}} +1)\ln{Z_\alpha}|_{\alpha=0}$.  Remarkably, this expression has the same form as thermodynamic entropy, $\sigma = (-\beta {\partial \over {\partial \beta}} + 1)\ln{Z}$, where $\beta$ is identified with the complete polar angle $2\pi$. 
\begin{figure}
\includegraphics[width=5cm]{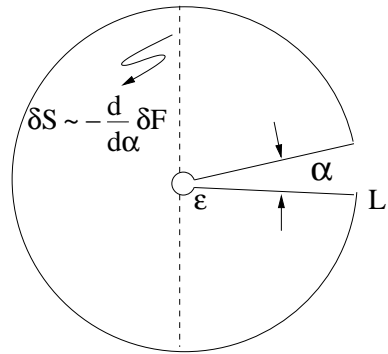}
\caption{\label{fig1} Conical geometry corresponding to path integral (\ref{partition}).  Deficit angle, $\alpha$, shown here is negative.  Dotted line is the (spacelike) impurity worldline.}
\end{figure}

The action appearing in (\ref{partition}) is taken to be the LL action, $S_0$, plus a weak impurity potential scattering term, $S_{\rm imp}$. To express the boundary conditions imposed in equation (\ref{partition}), the action is written in polar coordinates:
\begin{eqnarray}
\label{action}
S_\alpha &=& \frac{1}{g}\int_0^{2\pi}{d\theta\int_\epsilon^L{r  dr \half[(\partial_r \phi)^2 +\frac{1}{r^2}(\partial_\theta \phi)^2 ]}} \\
&+& y\sum_{\theta=\pm \frac{\pi}{2}}{\int{dr \cos{\sqrt{4\pi}\phi(r,\theta)}}}
\end{eqnarray}
Without the impurity, the computation of $\Sent$, at zero physical temperature, amounts to the computation of a conventional thermodynamic entropy from the free energy but on a conical manifold with the (imaginary) timelike angle variable having period $\beta \equiv 2\pi + \alpha$.  The corrections to the entropy from the boundary, $\delS$, then involve the {\sl finite temperature} ($\beta$) corrections of the impurity free energy, $\delta F$, to the bulk free energy, $F$; specifically, the variation of $\delta F$ with respect to $\alpha$ about $\beta=2\pi$.
\begin{equation}
\label{variation}
\delS = (-2\pi {d \over {d\alpha}} +1)(-\beta \delta F)|_{\alpha=0}
\end{equation}
\\

However, as depicted in figure 1, the impurity worldline has become spacelike.  This state of affairs is, in some sense (see figure 2), analogous to the computation of {\sl finite length} corrections to the {\sl pressure} in a Luttinger liquid due to a single impurity. The correction to the pressure is a variation of free energy with respect to length: $\delta p = -\partial \delta F/\partial L$, where $\delta F$ is the lowest order correction to the bulk free energy coming from the impurity.  In the finite size scaling regime in which $L \ll \beta$, $\delta F$, and therefore, $\delta p$, will depend algebraically on length, although they both vanish thermodynamically (i.e. $\delta F \sim L^{1-2g}$; $\delta p \sim L^{-2g}$).  In the entanglement entropy problem, the finite temperature scaling regime is reached when the quantization length scale exceeds the inverse temperature, $\beta = 2 \pi$.  It will be shown that this condition is $\ln{(L/\epsilon)} \gg \beta $, which can always be met.

To compute $\delta F$, the degrees of freedom in the bulk will be integrated out to produce an effective boundary sine-Gordon action for the "past" and "future" impurity fields at $\theta = \pm \pi/2$. There are two complicating characteristics: 1) The two impurity fields interact and 2) The impurity action lacks translation invariance which interferes with the standard momentum shell renormalization group scheme.

The free action (\ref{action}) is brought into a separable gaussian form with the "log-periodic" expansion:
\begin{equation}
\label{expansion}
\phi(r,\theta) = \frac{1}{\sqrt{2\pi P}} \sum_{k_m,\omega_n}{\sin{(k_m \ln{\frac{r}{\epsilon}})}e^{-i\omega_n\theta}\phi(k,\omega)}
\end{equation}
where we introduce Matsubara frequencies $\omega_n = \frac{2\pi n}{\beta} \equiv \frac{n}{1+\alpha/2\pi}$ that explicitly incorporate the deficit angle, wavenumber $k_m \equiv \frac{m\pi}{P}$ and $P \equiv \ln{\frac{L}{\epsilon}}$.  The spatial modes are made to vanish at $L$ and at a radial UV cut off, $\epsilon$.  The action now becomes:
\begin{equation}
\label{fourier_action}
S_0 = \frac{1}{g} \sum{\half(k_m^2 + \omega_n^2)|\phi(k\omega)|^2}
\end{equation}
Computation of $\Sent$ proceeds exactly as the standard thermodynamics computation except the final replacement of the sum over $k_m$ by an integral now includes a logarithmic density of states, $\ln{L/\epsilon}$.  Within the (Cardy-Callan-Wilzcek) conical geometry scheme, this is origin of the logarithmic entanglement entropy. 
\begin{figure}
\includegraphics[width=6cm]{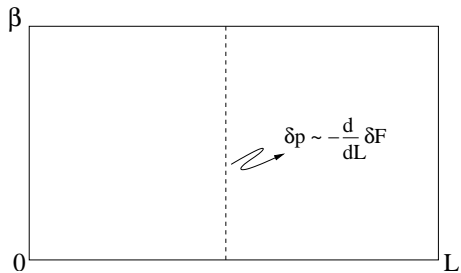}
\caption{\label{fig2} Analogous computation of finite $L$ correction to the pressure in the presence of an impurity (dotted line).}
\end{figure}

Following the standard procedure for deriving the effective impurity action, we introduce a pair of constraint fields, $\lambda_\pm(k)$, and impurity fields, $\phi_\pm(r)$, which correspond to the past and future branches of the impurity worldline. After the bulk fields $\phi(r\theta)$ are integrated out, the action for the constraint fields is:
\begin{eqnarray}
\nonumber
S_{\rm c} &=& \frac{g}{4} \sum_{k_m}{d^{-1}(k)
\left(  \begin{array}{c} \lambda_+\\ \lambda_- \end{array} \right)
\left(  \begin{array}{cc} 1 & \frac{1}{\cosh{\frac{\beta k_m}{2}}}\\ \frac{1}{\cosh{\frac{\beta k_m}{2}}} & 1 \end{array} \right)
\left(  \begin{array}{c} \lambda_+\\ \lambda_- \end{array} \right)}\\
&+& i \sum_{k_m; a+\pm}{\lambda_a (k) \phi_a (k)}
\end{eqnarray}
where $d(k) \equiv {k_m \tanh{\frac{\beta k_m}{2}}}$. We have also made use of the fact that $\phi^*(k\omega) = \phi(k -\omega)$, etc.  Finally, we integrate out the constraint fields and diagonalize the action using $\phi_\pm = \frac{1}{\sqrt{2}}(\phi \pm \bar{\phi})$. The final effective  impurity action is then:
\begin{eqnarray}
S &=& S_{0I} + S_{I} \\
\nonumber
&=&  \frac{1}{g} \sum_{k_m}{d(k_m)
\left(  \begin{array}{c} \phi \\ \bar{\phi} \end{array} \right)
\left(  \begin{array}{cc} \frac{1}{1+\Delta(k_m)} & 0\\ 0 & \frac{1}{1-\Delta(k_m)} \end{array} \right)
\left(  \begin{array}{c} \phi \\ \bar{\phi} \end{array} \right)}\\
\nonumber
&+&2y {\int_0^L{dr \cos{\sqrt{2\pi}\phi(r)}\cos{\sqrt{2\pi}\bar{\phi}(r)}}}
\end{eqnarray}
where $\Delta(k_m) \equiv 1/\cosh{\frac{\beta k_m}{2}}$.  This impurity model closely resembles the double barrier impurity model of \cite{KaneFisher} although the impurity worldlines are spacelike.  Similarly, the field $\bar{\phi}$, represents the number of particles "between" the two barriers and has an energy gap of $O(1/\beta)$.  There is no comparable "gate voltage" in this problem and thus $\<\bar{\phi}\> = 0$.  Only the mode $\phi$ is active and we reach the final effective low energy action:
\begin{equation}
\label{eff_action}
S = \frac{1}{g} \sum_{k_m}{d(k_m) \phi^2(k_m)} + y\int_\epsilon^L{dr \cos{\sqrt{4\pi}\phi(r)}}
\end{equation}

As expected, the finite time scale, $\beta$, appears in the effective spacelike model for the impurity. In an ordinary impurity model, renomalization group flow would terminate at the time scale $\beta$ and thus the algebraic behavior free energy could be directly determined.  However,
recalling the "log-periodic" substitution, equation (\ref{expansion}), this model no longer has translation invariance in $r$.  For this reason, it is not obvious how to implement a momentum shell
renormalization group calculation.  Instead, we directly compute the corrections to the bulk free energy arising from the impurity potential in (\ref{eff_action}) to lowest order in $y$:
\begin{equation}
\label{delta_f}
-\beta \delta F^{(2)} = \frac{1}{4} y^2  \int_\epsilon^L{dr_1 \int_\epsilon^L{dr_2 G(r_1,r_2) }}
\end{equation}
where 
\begin{equation}
\label{greens_fn}
G(r_1,r_2) \equiv \<e^{i\sqrt{4\pi}\phi(r_1)}  e^{-i\sqrt{4\pi}\phi(r_2)} \>_{S_{0I}}
\end{equation}
$G(r_1,r_2)$ is most easily calculated in terms of logarithmic variables, $p\equiv \ln{r/\epsilon}$:
\begin{equation}
\label{g(p)}
G(p_1,p_2) = \frac{ (\sinh{\frac{\pi(p_1+p_2)}{\beta}})^{2g} (\sinh{\frac{\pi(p_1-p_2)}{\beta}})^{-2g}}{(\sinh{\frac{2\pi p_1}{\beta}})^{g} (\sinh{\frac{2\pi p_2}{\beta}})^{g}}
\end{equation}
From the form of $G(p_1p_2)$, the free energy correction $\delta F^{(2)}$ will only be in the finite $\beta$ scaling range (and thus depend upon $\beta$ and the deficit angle, $\alpha$, as desired) if the characteristic logarithmic length scale, $p \gg \beta$.  Therefore, we arrive at the condition for 
finite $\beta$ scaling: $\ln{L/\epsilon} \gg \beta$.

Returning to radial variables, the "log-periodic" behavior is ultimately seen to leads to a large radial displacement Green's function that is algebraic in the {\sl ratio} of radii,
\begin{equation}
\label{g(r)}
G(r_1r_2) = \left| \left(\frac{r_1}{r_2}\right)^{\frac{1}{2+\alpha/\pi}} - \left(\frac{r_2}{r_1}\right)^{\frac{1}{2+\alpha/\pi}}\right|^{-2g}
\end{equation}
The IR divergent part of the correction $\delta F$ (\ref{delta_f}) comes from $r_2 \sim \epsilon$ and $r_1 \sim L$ where $G \sim (r_2/r_1)^{\frac{g}{1+\alpha/2\pi}} $ and thus we arrive at the leading order divergence in the impurity correction to the free energy:
\begin{eqnarray}
\label{delta_F}
\nonumber
-\beta \delta F^{(2)} &=& y^2 \epsilon^2 \frac{1}{1-(2\pi g/\beta)^2} \left[  \left(\frac{L}{\epsilon}\right)^{1-\frac{2\pi g}{\beta}} -1 \right]\\
&\sim& \left(\frac{L}{\epsilon}  \right)^{1-g+\frac{g\alpha}{2\pi}}
\end{eqnarray}
Finally, the boundary impurity correction, $\delS$, to the bulk entanglement entropy is found from the variation of $\delta F$ (equations (\ref{variation}) and (\ref{delta_F}), for $g \not= 1$):
\begin{equation}
\label{delta_S}
\frac{\delS}{y^2 \epsilon^2} = - \frac{g}{1-g^2}\left(\frac{L}{\epsilon}  \right)^{1-g}\ln{\frac{L}{\epsilon}}
- c(g)
\end{equation}
where $c(g)$ is a positive constant, independent of $L/\epsilon$.  For $g \rightarrow 1$,
\begin{equation}
\label{delta_S_g1}
\frac{\delS}{y^2 \epsilon^2} =  \frac{1}{(1+g)^2}\ln{\frac{L}{\epsilon}}\left(1-g\ln{\frac{L}{\epsilon}} \right)
\end{equation}

For $L/\epsilon \gg 1$, the entropy correction is strictly negative for all values of $g$.
The appearance of the inverse temperature, $\beta$, (or equivalently, the deficit angle $\alpha$) in the exponent of the free energy (\ref{delta_F}) is responsible for the negative value of the entropy, $\delS$, independent of $g$.  In a free massless gas, the free energy {\sl increases} to zero as the temperature goes to zero (or $\beta \rightarrow \infty$) and thus the variation (\ref{variation}) yields a positive entropy.  Here, the free energy decreases as $\beta \rightarrow 2\pi^-$.  

For  repulsive Luttinger liquids ($g<1$), $\delS$ diverges algebraically in $L/\epsilon$ suggesting that the bulk entanglement entropy, in the presence of an impurity, is zero for sufficiently large $L$.  Note, however, that for finite (bare) coupling $y$ and finite $L/\epsilon$, the entanglement entropy is nonzero, suggesting that high energy modes close to the boundary remain entangled until a critical coupling or length scale is reached.   For attractive Luttinger liquids ($g>1$), $\delS$ approaches a negative constant with algebraic corrections in $L/\epsilon$ that vanish as $L/\epsilon \rightarrow \infty$; therefore, an attractive LL remains maximally entangled.   For the noninteracting case ($g=1$), the correction  diverges as $\ln^2{L/\epsilon}$ suggesting that, even for the noninteracting fermions, the boundary impurity suppresses the entanglement entropy. 

Defining $y_0 \equiv y\epsilon$ as the dimensionless coupling on the scale of the lattice and $y_{\rm r} \equiv y_0 (L/\epsilon)^{1-g}$ as the renormalized coupling on the scale, $L$ (using the RG results of \cite{KaneFisher}), $\delS = -y_0 y_{\rm r} \ln{\frac{L}{\epsilon}}$.  Our perturbative results may then be improved by the renormalization group and written in the following suggestive way:
\begin{equation}
\label{improved_S}
\Sent = \left(\frac{1}{3} - y_0 y_{\rm r}\right)\ln{\frac{L}{\epsilon}}
\end{equation}
for $g<1$ (using $c=1$ for a spinless LL).  In \cite{Vidal}, entanglement loss was suggested as a fundamental feature of the irreversibility of renormalization group flow.  This intriguing proposal followed along the lines of the "c-theorem" of conformal field theory \cite{Zamolodchikov}.  Refael and Moore \cite{Refael} have recently computed the entanglement entropy for strongly disordered spin chains and found that $\Sent$ remains logarithmic, but with a universal prefactor, $\tilde{c}$ which is smaller than the central charge of the clean system, consistent with this proposal. In the zero physical temperature KF impurity model, RG flow is accessed through $L/\epsilon \rightarrow \infty$ (which also drives the system further into the finite $\beta$ scaling limit on which our results are based.)  Thus, noting the sign change in the $g$-dependent prefactor of (\ref{delta_S}), $\delS$ strictly decreases for all $g$ along the renormalization group trajectory set by $y_0$ and increasing $L/\epsilon$.  

Lastly, we find it significant that the logarithmic behavior of the entropy (eqns. (\ref{delta_S}), (\ref{delta_S_g1}) and (\ref{improved_S})) is now a symptom of an IR divergence in the boundary theory, rather than the logarithmic divergence of the wavenumber eigenvalues ($k_m^{-1} \propto \ln{L/\epsilon}$) responsible for the bulk entanglement entropy within the Cardy-Callan-Wilczek construction.

It should be noted that these results pertain to a LL {\sl bisected} by an impurity; it  is unknown whether this result remains true for a region of length $L$ imbedded in an infinite LL. (Recently, Calabrese and Cardy have given generalizations to the bulk entanglement entropy for unequal partitions of $1+1$ dimensional CFTs \cite{Cardy_rev}.)  Furthermore equation (\ref{improved_S}) does not reflect the contribution to the entropy from the open boundaries at $\pm L$ implicit in our construction \cite{Cardy_rev}.

The bosonization scheme presented here may likely be applied to Kondo type impurities as well as spin chains with comparable boundary impurities. Entanglement of an impurity spin with conduction channels plays a crucial role in more complicated boundary problems such as the two-channel Kondo effect. Furthermore, the predictions for the behavior of the entanglement entropy in the presence of an impurity might be checked by density matrix renormalization group (DMRG) methods, since DMRG explicitly follows the flow of eigenvalues of the reduced density matrix.  

The author wishes to thank Gil Refael and Oleg Starykh for very useful discussions, Christian Hilaire and Alex Zaharakis for {\sl Mathematica} support, and the Kavli Institute for Theoretical Physics at UCSB where this work was completed; this research was supported in part by the National Science Foundation under Grant No. PHY99-07949.


\begin{references}
\bibitem{basis_explanation} At scale $n$, the wavenumber $k=2^nk_{\rm min}$ contributes $O(kL)$ spatial modes, only one of which straddles the boundary and is entangled.  However, the maximum number of "binary" scales is $n_{\rm max} \sim \log{k_{\rm max}/k_{\rm min}}$, where $k_{\rm max}$ is the UV cut off $\epsilon^{-1}$.
\end{references}
\end{document}